\documentclass[conference]{IEEEtran}
\IEEEoverridecommandlockouts
\usepackage{cite}
\usepackage{amsmath,amssymb,amsfonts}
\usepackage{comment}
\usepackage{algpseudocode}
\usepackage{hyperref}
\usepackage{algorithm}
\usepackage{tikz}
\usepackage{textcomp}
\usepackage{tabularx}
\usepackage{xcolor}
\usepackage{url}
\usepackage{csquotes}
\usepackage{enumerate}
\usepackage{listings, lstautogobble}
\usepackage[utf8]{inputenc}
\usepackage{graphicx}
\usepackage[caption=false,font=footnotesize]{subfig}

\def\BibTeX{{\rm B\kern-.05em{\sc i\kern-.025em b}\kern-.08em
    T\kern-.1667em\lower.7ex\hbox{E}\kern-.125emX}}

\def\sysname{spaPS}

\newcommand\copyrighttext{%
  \footnotesize This is the author’s version of the work. It is posted here for your personal use. Not for redistribution.
A revised and more concise version of this work is available in proceedings of the DISCOLI workshop of the 22nd IEEE International Conference on Distributed Computing in Smart Systems and the Internet of Things (DCOSS-IoT), 2026.}
\newcommand\copyrightnotice{%
\begin{tikzpicture}[remember picture,overlay]
\node[anchor=south,yshift=10pt] at (current page.south) {\fbox{\parbox{\dimexpr\textwidth-\fboxsep-\fboxrule\relax}{\copyrighttext}}};
\end{tikzpicture}%
}

\begin{document}

\title{
 A Spatially-Aware Publish–Subscribe Middleware for IoT Applications
}

\author{
\IEEEauthorblockN{
   Philipp Ungrund\IEEEauthorrefmark{1}, 
    Kurt Rothermel\IEEEauthorrefmark{2}
    Sukanya Bhowmik\IEEEauthorrefmark{1}, 
  }
  \IEEEauthorrefmark{1}University of Potsdam, Germany,
  \IEEEauthorrefmark{2}University of Stuttgart, Germany \\ Email:
  \IEEEauthorrefmark{1}lastname@uni-potsdam.de,
  \IEEEauthorrefmark{2}kurt.rothermel@ipvs.uni-stuttgart.de 
}

\maketitle
\thispagestyle{plain}
\pagestyle{plain}

\begin{abstract}
Spatial and proximity awareness are critical enablers for efficient communication in Cyber-Physical Systems (CPS) and the Internet of Things (IoT). However, the lack of suitable middleware support and standardized mechanisms for spatial awareness significantly limits the pervasiveness of location-dependent applications. In this paper, we present a novel approach that extends topic-based publish–subscribe, a dominant messaging model in this domain, with integrated spatial filtering and routing capabilities. Our approach leverages a highly flexible geometric method for describing spatial filters, enabling expressive and efficient spatial subscriptions. A geometric world model operates together with an enhanced topic-based broker to realize spatially constrained message dissemination. Importantly, this functionality is achieved using the MQTT5 protocol without violating the standard or requiring any modifications to existing client implementations. Experimental results based on both synthetic and real-world data demonstrate strong system efficiency, minimal computational overhead at the broker, and the practical feasibility of deploying the approach on existing IoT devices.

\end{abstract}


\copyrightnotice

\section{Introduction} \label{sec:introduction}

The Internet of Things (IoT) is a significant technology that profoundly impacts everyday life, industries, and the economy. 
IoT technology focuses on connecting devices and enabling data exchange in networked environments.  It is therefore a building block for Cyber-Physical Systems (CPS), which emphasize the close integration of computing, networking, and physical processes to monitor and control real-world assets and processes. Digital twins leverage CPS models and IoT data to enable real-time monitoring, simulation, and optimization of their physical counterparts. 

The publish-subscribe (pub/sub) communication par\-a\-digm has been identified as a basic building block of the IoT. Pub\-lishers send messages without knowing the recipients, and sub\-scribers register for the messages they are interested in without knowing the senders. In topic-based pub/sub, messages are associated with topics, and subscribers register for topics of interest. This communication model allows many-to-many communication, decoupling senders from receivers for better scalability and flexibility. The primary standard for pub/sub communication in the IoT is MQTT \cite{4554519}, 
alongside protocols like AMQP~\cite{Vinoski2006}, CoAP~\cite{shelby2014}, etc.

A distinctive feature of CPS is that they maintain digital representations of their physical objects.
These digital entities keep state data, which reflects the physical, operational and contextual status of their physical counterparts. This  data forms the basis for monitoring, controlling, and optimizing physical entities and real-world processes. Digital entities are also referred to as resources \cite{OCF2025}, things \cite{WoT2023}, objects \cite{OPC2025}, or digital twins \cite{ISO2023} elsewhere.

The state of an entity often depend on the states of other physical entities. Therefore, digital entities need to be informed about state changes of the physical entities they depend on, their group of interest.
For example, a digital entity representing a particular intersection wants to receive traffic flow data for all adjacent road segments to determine the intersection's current traffic load. Physical entities, like roads and intersections, are situated in space. Therefore, the spatial relationships between these entities provide a powerful means of selecting groups of interest. These relationships define how physical entities interact geometrically; for example, entities may contain, touch, or intersect each other. In our example above, the spatial relation "touches" would be used to determine which road segments are of interest.

Few pub/sub systems support spatial relationships as a message filtering concept. Location-based pub/sub systems (e.g.,\cite{Chapuis2017,Fisher2019,Guo2015,Hasenburg2020,Ihirwe2021}) support geofencing, a concept that allows subscribers to define an area of interest for each topic. Messages are only forwarded to consumers if they were published in the corresponding area of interest. However, these schemes are limited to a single spatial relation: containment. Furthermore, the location of publishers is used for filtering messages rather than that of the owner of the transferred state. This requires the state owner to be physically co-located with the publisher, which severely limits the placement of publishers.

In this paper, we propose \sysname, a novel spatial pub/sub middleware tailored to the specific needs of CPS. With \sysname, we expand the publish/subscribe paradigm to include physical entities, thereby accommodating the physical aspect of CPS. The digital counterparts of physical objects, which can act as publishers and consumers, represent the virtual part of CPS. Our system supports topic-based message filtering, although the proposed spatial extensions could also be combined with content-based filtering.

Each published message is assigned a topic, which indicates the type of state data included in the message. Additionally, it is associated with the physical entity whose state is transferred in the message. Consumers subscribe to topics and groups of interest, called \textit{neighborhoods}, inspired by Tobler's first law of geography \cite{Tobler1970}, which states that nearby things are more relevant than distant ones. In particular, each subscription specifies topics and a neighborhood, indicating the physical entities whose state data the subscriber wants to receive. 


\sysname's architecture leverages the inherent customizability of MQTT version 5 and completely conforms to the standard. With this approach, all additional complexity is confined to the broker while client implementations can stay virtually unchanged. Thus, publishers and subscribers can continue to run on resource-limited devices and the protocol's lightweight character is kept. For the broker's implementation, we enhance ActiveMQ Artemis as an open source, multi-core, and highly performant basis \cite{mishra2021stress}. 

The contributions of this paper are as follows: 
\begin{itemize}
    \item An expressive scheme for specifying neighborhoods based on spatial relationships, which are the foundation of subscriptions.  
    \item  An architecture that incorporates spatial filtering on these neighborhoods into classic topic-based pub/sub systems while adhering to the MQTT standard.
    \item An implementation of this architecture on the ActiveMQ Artemis broker that can be used as a drop-in replacement.
    \item Extensive evaluations of \sysname~that demonstrate the practical feasibility and efficiency of the novel middleware.
\end{itemize}

\section{Related Work} \label{sec:related}

Pub/sub has been widely adopted by modern geographically distributed applications to perform event-based communication~\cite{10.1145/857076.857078}. Most efforts have focused on achieving high throughput and low end-to-end latency in scalable pub/sub systems ~\cite{Li:bluedove, TariqKBR14, BhowmikTGSR18, BhowmikTHR16, Barazzutti:streamhub, Jokela2009LIPSIN}. While systems such as BlueDove~\cite{Li:bluedove} and StreamHub~\cite{Barazzutti:streamhub} target parallelism of the event filtering process on multiple servers, those such as PLEROMA~\cite{TariqKBR14, BhowmikTKD0R17} and LIPSIN\cite{Jokela2009LIPSIN} perform event filtering on the network layer to achieve line-rate performance. However, given the emphasis of these works, integrating spatial awareness into them is rather challenging. This has led to the emergence of specialized location-based pub/sub systems.


With location-based pub/sub, subscribers can use geofencing to define areas of interest (e.g., \cite{Chapuis2017,Fisher2019,Guo2015,Hasenburg2020,Ihirwe2021}. Each message is associated with the publisher's location. Messages can only be forwarded to subscribers if the publisher's location falls within one of their areas of interest. All of these systems are based on a simple location model that includes the locations (latitude, longitude) of producers and consumers, as well as a collection of geofences of various shapes, like circles or polygons. However, this simplicity comes with limitations. First, "containment" is the only spatial relation exploited. Second, since messages are associated with the publisher's location rather than the physical entity whose state is included in the message, publishers must be co-located with the corresponding physical entities. This severely limits publisher placement. \sysname's richer world model allows for decoupling physical and digital entities, with the latter being producers and consumers that can be placed anywhere in the IoT infrastructure, including on mobile devices. 

For scalability reasons, some location-based pub/sub systems allow publishers to associate messages with an area of relevance (e.g., \cite{Hasenburg2020,Montori2022}), meaning a consumer can only receive a message if it is currently located in the corresponding area.  While neighborhoods could be used to define groups of relevance, we will not expand on this issue further.

 Like \sysname, some location-based pub/sub systems conform to the MQTT standard (e.g.,\cite{Montori2022,Fisher2019}). However, it is important to note that \sysname's concepts and architecture do not depend on any MQTT features. Additionally, one of these systems, LA-MQTT~\cite{Montori2022}, performs location filtering and routing within a dedicated pub/sub client, leading to each publication having to traverse the broker twice. This decreases efficiency and is a conceptual disadvantage in scalability. 

A proximity-based service discovery scheme for the IoT is presented in \cite {Rothermel2025}. Although this scheme also uses spatial relations to define service proximity, the neighborhood concept proposed in this paper is tailored to the specific needs of pub/sub communication. Furthermore, the spatial extensions for service discovery differ substantially from those for pub/sub.


\sysname~is based on a geometric model that encompasses the relevant physical entities. 
Appropriate models are widely available, e.g., Geographic information system (GIS) models like OpenStreetMap, NASA EarthData. 
A vast amount of publicly available geospatial data comes from sources such as the USGS Earth Resources Observation and Science, NASA Earthdata, as well as from crowd-sourced initiatives like OpenStreetMap. 
Furthermore, Building Information Modeling (BIM) \cite{Eastman2011} is widely adopted in the architecture, engineering, and construction industry.


\section {System Model}
\begin{figure} [htb]
	\centering
	\includegraphics [width=0.7\linewidth]{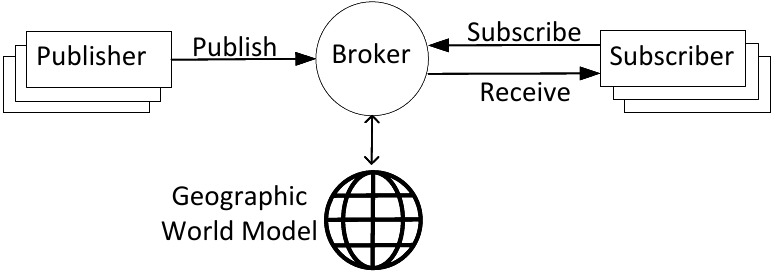}
	\centering
	\caption{System Components}
	\label{systemmodel}
\end{figure}
In this section, the major components and concepts of \sysname~will be introduced. As illustrated, in Fig.~\ref{systemmodel} the system model comprises the following components: publishers, subscribers, the broker, and the geometric world model. In the following, we will describe these components in detail.  

\subsubsection{Publishers, Subscribers, and Broker}

Digital entities can publish their state data for consumption by other digital entities or application components.
In \sysname, each published message is associated with a topic and a state owner. The topic defines the type of state data that the message transfers, while the state owner is the physical entity to which the state belongs.

Since messages are also associated with physical entities in \sysname, subscribers can filter messages by both topic and state owner. A subscriber registers for a set of topics and a neighborhood, specifying the state owners in which it is interested.
For instance, a traffic management entity controlling the traffic lights of an intersection, say \textit{I}
would subscribe to topic \(/trafficFlow/\#\) and  a neighborhood  encompassing all road segments that are adjacent to  \textit{I}. Consequently, this entity would receive all traffic flow data directly affecting the traffic load of \textit{I}. As mentioned above, 
neighborhoods are specified based on the spatial relations between physical entities. We will propose an expressive scheme for specifying neighborhoods in Sec.~\ref{secNeighborhood}.

The message broker functions as a central intermediary, thereby decoupling publishers from subscribers. The broker is responsible for receiving messages, filtering them based on topics and neighborhoods, and subsequently routing them to all interested subscribers. The determination of neighborhoods necessitates spatial analysis, for which the broker employs a geometric model encompassing the geometry of the relevant physical entities as well as their spatial relations.
This model, discussed next, is referred to as the Geometric World Model. 


\subsubsection {Geometric World Model}
\label{WorldModel}

The Geometric World Model (in short World Model) includes a geometric representation of all physical entities relevant to the application. These entities can be visible material objects, such as buildings or road segments, or immaterial objects, such as traffic zones or national borders. 

A physical entity is represented by three attributes:
\begin{itemize}
	\item \textit{Identifier} (denoted by \emph{peid}), which is unique among all physical entities in the model.
	\item \textit{Geometry} (\emph{geom}), which defines the entity's position, orientation, size and shape in a coordinate system. 
    \item \textit{Category} (\emph{Cat}) to which the entity belongs. Categories typically group entities based on their purpose or type. 
\end{itemize}

Like topics, categories can be organized hierarchically. 
For instance, the category \(road/major/motorway \) would include the motorway segments in the world model.
The World Model also  contains spatial relationships between the physical entities.
A great variety of models for topological, metric and directional relations has been proposed in literature~\cite{Chen2013}. In our world model, we will use the relations standardized by the OGC Simple Feature Access \cite{OGC2010}. Some of these relations, along with their abbreviations and descriptions, are shown in Table \ref{OGCRelations}. The last relation listed in this table is a metric relation, whereas all the others are topological relations. 

 \begin {table}[htb] 
 \caption {Spatial Relations Standardized by OGC Simple Feature Access  \cite{OGC2010}}
 \label {OGCRelations}
 \begin{tabular}{lc>{\raggedright\arraybackslash}p{4.6cm}}
 	\hline
 	Relation & Abbrev. & Description \\
 	\hline
 	Equals(a,b) & $E$  & geometries a and b are topological equivalent. \\ 
 	Intersects(a,b)& $I$ & a and b have at least one point in common. \\
 	Disjoint(a,b) & $D$ &  a and b have no point in common. This is the logical negation of Intersects(a,b). \\
 	Touches(a,b) & $T$ & a and b have at least one boundary point in common, but no interior points. \\
 	Contains(a,b) & $C$ &  b lies a and their interiors intersect.\\
 	\hline
 	DWithin(a,b,d) & $DWd$ & The distance between geometries a and b is less or equal d. \\
 	\hline
 \end{tabular}
 \vspace{-5pt}
 \end{table}

The OGC Simple Feature Access also defines a set of SQL routines for spatial analysis. In particular, for each of the depicted relationships, there exists a Boolean function that returns true if two geometries relate to each other and false otherwise. All major DBMS vendors support these routines.

We will assume that the world model is quasi-static, similar to maps. This means that changes will result in new model versions.
Consequently, the world model does not support mobile physical entities, such as vehicles or people. However, note that this restriction applies only to modeled physical entities, whereas subscribers and publishers may be mobile. Supporting mobile physical entities is subject to future research.
\section{Specifying Neighborhoods}
\label{secNeighborhood}

Neighborhoods are the core concept of \sysname. In this section, we will introduce an expressive method for specifying neighborhoods based on spatial relations.

As mentioned above, the simplest form of a neighborhood specification consists of three attributes: a reference entity, a category, and a spatial relation. This specification can be extended in several ways to increase its expressiveness. First, instead of a single reference and category, we can allow for sets of references and categories. Secondly, rather than a single spatial relation, we can allow multiple relations connected by the logical operators "AND," "OR," and "NOT." These extensions lead to the following neighborhood specification:
\begin{equation} \label{singleStageNH}
	Name: (Refs,Cats,Cond)
\end{equation} 
Each neighborhood definition consists of a neighborhood name and a neighborhood descriptor, including three attributes:
\(Refs\) is a non-empty set of references, \(Cats\) denotes a non-empty set of categories, and \(Cond\) represents a single spatial relation or several logically connected spatial relations.
For example, suppose that the entity responsible for traffic control in zone \(z\)  is interested in the real-time traffic flow data of primary road segments in that zone. The corresponding neighborhood can be specified as follows: \(N_1:(\{z\}, road/major/primary, C)\), where \(C\) is the abbreviation for relation \(Contains\) listed in Table \ref{OGCRelations}. Now, assume that this entity only wants traffic data of its border motorway segments, i.e., the segments in \textit{z} that touch \textit{z}'s border: \(N_2:(\{z\},\{road/major/motorway/\},C \wedge T)\). If the entity were responsible for zones \(z1\) and \(z2\) and  were interested in all major roads in these zones, the corresponding neighborhood would be \(N_3:(\{z1,z2\},\{road/major/\#\},C)\). Note  wildcard \(\# \) in the category pattern.

To precisely define the semantics of the above neighborhood notation, we introduce two functions. Boolean function \(Related(e1,e2, Cond)\) checks whether  entities \(e1\) and \(e2\) are spatially related according to condition \(Cond\). It returns true if they are related and false otherwise. Note that this function encapsulates the logic for the spatial analysis.


Function \(Select\) determines the neighborhood specified by parameters \(Refs\), \(Cats\) and \(Cond\). Let \(Entities(Cats)\) denote the set of entities belonging to the categories specified by \(Cats\). 
\vspace{-6pt}
\begin{multline}
\label{Select}
Select(Refs,Cats,Cond) = \{ e \in Entities(Cats)-Ref: \\
\exists_{r \in Refs} Related(r, e, Cond)  \}
\end{multline} 
This function determines those entities in \(Entities(Cat)-Ref\) that are spatially related to at least one entity in \(Refs\) according to condition \(Cond\). For example, neighborhood  \(N_4:(\{seg1,seg2\},\{road/\#\},WD1000)\)  selects all road segments within 1000 meters of segments \(seg1\) or \(seg2\). However, \(N_4\)  does not include \(seg1\) or \(seg2\), even if the two references are 1000 meters or less apart.

To further increase expressiveness, we introduce multi-stage neighborhoods. We will refer to the neighborhood introduced above as a single-stage neighborhood because it is defined by a single  descriptor. In contrast, multi-stage neighborhoods consist of a sequence of chained stages, each of which is specified by an individual stage descriptor. The selection process begins with the first stage and then moves on to subsequent stages. Each stage selects a set of neighboring entities that may serve as input for subsequent stages. While the first stage uses the reference set supplied by the application, each subsequent stage determines its reference set based on neighbors selected in previous stages. Specifically, a stage can adopt the neighbors selected by any subset of its preceding stages. 

Assume that an entity responsible for a city's parking management   wants to track the occupancy status of parking facilities that are adjacent to \(z\)'s border motorway segments. This neighborhood requires two stages: The first stage yields the border motorway segments of \(z\) (see \(N_2\) above), which then are adopted as reference set of the second stage. Finally, the second stage selects parking spaces adjacent to at least one entity in its reference set.
We will use the following notation to specify a neighborhood with k stages: 
\begin{equation} \label{multiStageNH}
\begin{split}
Name:~&(Refs,Cats_1,Cond_1)~\bullet \\
&... \\
&(RefStages_k,Cat_k,Cond_k)
\end{split}	
\end{equation} 
\vspace{-6pt}

A multi-stage neighborhood has a name and consists of a sequence of stage descriptors.
The descriptor for the first stage consists of the attributes of a single-stage neighborhood as introduced above. The stage descriptors for subsequent stages differ from this structure only in their first parameter \(RefStages_i\). While parameter \(Refs\) of stage 1 passes the set of references supplied by the application, \(RefStages_i\) specifies, how the reference set is to be constructed from neighbors selected by the stages preceding stage \(i\). 
Let the set of stages preceding \(i\) be represented by \(\{1,2,..,i-1\}\). Then \(RefStages_i\) can be any non-empty subset of this set. Each stage identified by \(RefStages_i\) contributes neighbors to the set of references of stage \(i\). We will use shorthand \(Prev\) and \(Anc\) to indicate that \(RefStages_i\) is set to  \(\{1,2,...,i-1\}\) and  \(\{i-1\}\), respectively.

The following equations define how the neighbors of a k-stages neighborhood named \(N\) are determined. The neighbors selected in stage \(i\) will be denoted \(N[i]\): 
\begin{equation} \label{multiStageNH}
\begin{split}
N[1] &= Select(Ref_1,Cat_1,Cond_1),  \\
Ref[i] &= \bigcup_{j \in RefStages_i} N[j], \\
N[i] &= Select(Ref[i],Cat_i,Cond_i),  \\
&(2 \le i \le k) \\
\end{split}	
\end{equation} 

The first equation defines N[1], the neighbors selected in stage 1. The second equation shows how the reference set of stage \(i\) is built based on parameter \(RefStages_i\) and the neighbors of preceding stages. Finally, the third equation defines how the neighbors of stage \(i\) are selected based on this reference set. 

To illustrate the effect of the various settings of parameter \(RefStages\), we will examine the neighborhoods in Table \ref{ExampleNH}.
 \begin {table}[htb] 
 \caption {Example neighborhoods} \label{ExampleNH}
 \begin{tabularx}{\columnwidth}{>{\raggedright\arraybackslash}X}
 	\hline

\(N_5:(\{z\},\{road/\#\},C)\bullet 3(Prev,\{road/\#\},T)\) \\


\(N_6:(\{z\},\{road/\#\},C)\bullet 3(Prev,\{road/\#\},T)
\allowbreak\text{\hspace{11em}}\bullet (\{2,3,4\},\{transport/parking\},T\) \\


 	\hline
 \end{tabularx}
\end{table}

For neighborhood \(N_5\), we assume that the traffic control entity subscribes to traffic data from the road segments in zone \(z\). To take also into account the traffic situation in neighboring zones, it additionally needs to receive traffic data from road segments within 3 hops of \(z\).  For stages 2 through 4 the \(RefStages\) parameter is set to \(Prev\), which ensures that the neighbor sets selected by the 4 stages are mutually disjoint. \(N_5[1]\) contains all segments contained in \(z\), and \(N_5[i]\) \((2\le i \le 4)\) include all segments at a distance of \((i-1)\) hops from \(z\). 
For neighborhood \(N_6\), we assume a parking management entity that tracks the occupancy status of parking facilities that are accessible via the road segments identified by \(N_5\). Neighborhood \(N_6\) has an additional 5th  stage that specifies neighbors in the \(transport/parking\) category. 
In neighborhood \(N_6\), only stages 2 through 4 contribute to the reference set of stage 5. Therefore, \(N_6[5]\) includes only  parking spaces accessible via road segments outside \textit{z} and reachable within three hops from \textit{z}.

Finally, we will introduce the concept of a \textbf{\textit{stage filter}}. So far, a k-stage neighborhood includes all neighbors \(N(i)\) for \(1 \le i \le k\). However, a subscriber might be only interested in neighbors of a subset of stages.
In the example above, the parking management entity is only interested in the parking facilities selected in the final stage of neighborhood \(N_5\). 
To enable subscribers to choose which stages are included in the final neighborhood, we attach parameter \(Visible\), which is a Boolean vector with dimension \(k\) for a \(k\)-stage neighborhood. Entities selected in stage \(j\) will only be included if \(Visible[j]\) is set to true. The final neighborhood is the union of the neighbors selected by visible stages of \(N\). 



\section{The \sysname~Middleware} \label{sec:middleware}
\subsection{System Architecture}

\begin{figure}[h!]
    \centering
    \includegraphics[width=0.79\linewidth]{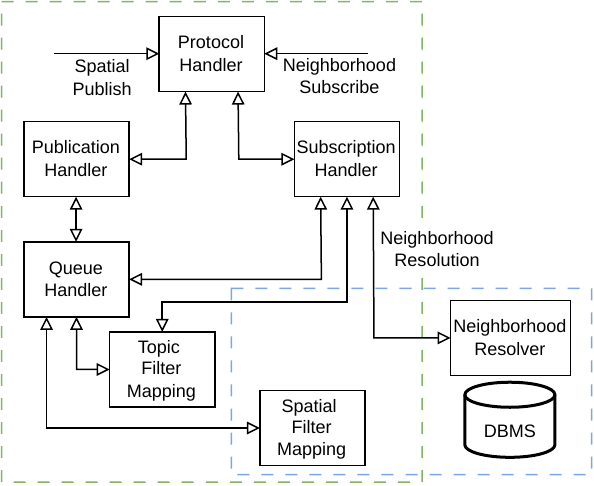}
    \caption{Broker Architecture}
    \label{fig:system-architecture}
\end{figure}

The overall system architecture of \sysname~ follows the system model in Figure~\ref{systemmodel}. The system components of this architecture are comprised of the \emph{spatial broker} - an enhanced topic-based pub-sub broker that integrates with a spatial database - and \emph{clients}, which can assume publishing and subscribing roles. The spatial database holds the world model and is queried by the \sysname~broker for neighborhood resolution. The details of our broker architecture is depicted in Figure~\ref{fig:system-architecture}.

As mentioned above, a neighborhood subscription consists of a topic and a neighborhood. We refer to the action of issuing such a subscription as a \emph{neighborhood subscribe}. Issuance of a neighborhood subscribe within the broker triggers a corresponding service provided by the spatial database called \emph{neighborhood resolution}. This neighborhood resolution produces a \textit{mapping} between the specific subscription and the relevant physical entity identifiers (\emph{peid}), which we refer to as \textit{spatial filters}. 
Conversely, we refer to the action of sending a publication that includes a \emph{peid} as a \emph{spatial publish}. 

Internally, the \sysname~broker receives publications and subscriptions via a protocol handler, which interfaces with a (i) subscription manager and a (ii) publication manager. For each subscription of a subscriber, the broker maintains a message queue to deliver publications. The subscription manager maintains the state of the topic filter mapping and the spatial filter mapping, and updates a queue manager that maintains the aforementioned queues. It also employs the neighborhood resolver to add new spatial mappings to the spatial filter mapping in response to neighborhood subscriptions. Conversely, the publication manager leverages the queue manager, topic filter mapping, and spatial filter mapping to filter publications and route them to the appropriate subscription message queues.


\subsection{Subscription Handling}
For each incoming subscription four steps are performed: (i) a new message queue is created and associated with the subscription, (ii) the subscription's topic and identifier are added to the topic filter mapping, (iii) the neighborhood is resolved via the spatial DBMS with a neighborhood query, and finally (iv) the neighborhood's resolved set of \emph{peid}s is added to the spatial filter mapping. Next, we discuss the last two steps in detail.

\subsubsection{Processing Neighborhood Queries}
The following outlines how neighborhood resolution queries are processed on the DBMS site. 
To improve readability, we will use pseudocode rather than SQL to sketch the major stored procedure that resolves a given neighborhood into a set of neighbors. 

The used DBMS tables are depicted in Listing~\ref{Tables}. 
Table \texttt{PhysEntities} contains a single row for each physical entity, which comprises the identifier and geometry of the corresponding entity. Note that this information is available in virtually all geometric models; hence the content of this table can be extracted from existing data sets. Table \texttt{catMembers} implements a many-to-many mapping between categories and physical entities. View \texttt{catLayers} adds geometry information to each pair (\texttt{peid}, \texttt{cat}), i.e., this table represents the category layers of our world model. \texttt{NTable} is a temporary table that is used to compute the members of a neighborhood. When the computation is finished, it contains the final neighborhood, where each member is identified by \texttt{peid}, and \texttt{geom} and  \texttt{stageNo} specify the entity's geometry and the stage in which it was selected, respectively.

\lstdefinestyle{mySQLStyle}{
    language=SQL,
    basicstyle=\footnotesize\ttfamily,
    morekeywords={TYPE, FUNCTION, RETURNS, RETURN, DECLARE, BEGIN, END, LOOP, FOR, IN, IF, THEN, ARRAY, MATERIALIZED, VIEW, BOOLEAN},
    keywordstyle=\color{black}\bfseries, 
    ndkeywordstyle=\color{darkgray}\bfseries,
    identifierstyle=\color{black},
    commentstyle=\color{gray}\itshape,
    stringstyle=\color{purple},
    columns=flexible,
    escapeinside={(*}{*)},
    numbers=left,
    numberstyle=\tiny\color{gray},
    stepnumber=1,
    numbersep=5pt,
    breaklines=true,
    tabsize=1,
    xleftmargin=1.5em,
    showstringspaces=false,
    captionpos=b
}

\begin{lstlisting}[style=mySQLStyle, caption={DBMS Tables}, label=Tables, float=htb]
TABLE physEntities (peId physEntityId,  geom geometry)
TABLE  catMembers (cat Category, peid physEntityId)
MATERIALIZED VIEW catLayers AS
	 SELECT c.peId, c.cat, p.geom 
	 FROM catMembers AS c JOIN physEntities AS p ON c.peId=p.peId
TEMPORARY TABLE NTable (stageNo int, peId physEntityId, geom geometry)
\end{lstlisting}

The algorithm for selecting neighborhoods is outlined in Listing \ref{StoredProc}. It  shows the major data types and the stored function that computes the neighborhood for a given neighborhood descriptor.  A neighborhood descriptor is passed in an array of type \verb|nhoodDes|, which includes a stage descriptor for each specified stage. A stage descriptor of type \verb|stageDes|, specifies the neighbors to be selected in the corresponding stage. It consists of the components \verb|refStages|, \verb|cats|, and \verb|sCond|. The component \verb|refStages| lists the previous stages whose selected neighbors are to be merged to constitute the reference set of the current stage. Components \verb|cats| and \verb|sCond| define the desired neighbor categories and the spatial condition that a member must fulfill, respectively. Finally, the entity descriptor \verb|entityDes| comprises the identifier and geometry of the corresponding physical entity.
\begin{lstlisting}[style=mySQLStyle, caption={Algorithm for Selecting Neighborhoods}, label=StoredProc, float=htb]
TYPE stageDes (refStages int, cats Category[], sCond spatialCondition)       
TYPE nhoodDes stageDes[];
TYPE entityDes (peid physEntityId, geom geometry)

FUNCTION selectNeighborhood (refs physEntityId[], nd nhoodDes, visible BOOLEAN[])	
  RETURNS physEntityId[];
  DECLARE 
    refEntities entityDes[];
    stageNeighbors entityDes[];
  BEGIN
    CREATE TEMPORARY TABLE NTable; 
    refEntities := GetGeometry(refs);
    InsertIntoNTable (0,refEntities);
    nd[1].refstages := ARRAY[0];
    FOR i IN 1..array_upper(nd,1) LOOP
      refEntities := mergeRefEntities(nd[i].refStages);
      stageNeighbors:=selectStageNeigh(refEntities, sd[i].cats, sd[i].sCond);	
      insertIntoNTable(i,stageNeighbors);
    END LOOP;
    RETURN filterNTable(visible); 
  END;
\end{lstlisting}
Function \verb|selectNeighborhood| selects the neighborhood specified by descriptor \verb|nd| and references \verb|refs|. Parameter \verb|visible| defines which stages will be visible in the final result. First, temporary table \verb|NTable| is created, which is used to record the selected neighbors during stage processing. Next, function \verb|GetGeometry| retrieves the geometry of each physical entity identified by \verb|refs| from table \verb|catLayers|. The resulting entity descriptors are inserted in \verb|NTable| as stage 0 neighbors. By setting \verb|refStages| of the stage 1 descriptor to refer to stage 0, allows stage 1 to be treated as any subsequent stage. 

The body of the loop is executed once for each specified stage. 
First, function \verb|MergeRefEntities| delivers the reference entities for the current stage, say \textit{i}, by merging the stage neighbors of the stages identified by \verb|refStages| in \textit{i}'s descriptor. This function operates on table \verb|NTable|, which records all neighbors selected in stages preceding \textit{i}. Next, function \verb|selectStageNeigh| computes the stage \textit{i} neighbors based on \verb|refStages| as well as the categories and spatial conditions specified for stage \textit{i}. This function operates on table \verb|catLayers| and uses the spatial predicates provided by the DBMS \cite{OGC2010} to analyze which entities fulfill the spatial conditions. Processing of stage \textit{i} ends with inserting the selected stage \textit{i} neighbors into table \verb|NTable|. After processing all stages, function \verb|filterNTable| returns the identities of neighbors belonging to visible stages. 


\subsubsection{Spatial Filter Mapping}
As explained above, the neighborhood resolution returns a set of \emph{peid}s for each subscription, resulting in an update to a spatial filter map. 
This map contains a mapping of unique subscription identifiers to a set of \emph{peid}s.


For this, the architecture distinguishes between two main conceptual implementation variants that differ in where and how the mapping is stored. The first mode stores the mapping as a table within the DBMS and makes those tables available to the broker through stored procedures. 
The second mode stores the spatial filter mapping within the broker's main memory.

The table of the in-database variant has two columns: one for the subscription identifiers and another for the \emph{peid}s. Both columns have a B-tree index on them, resulting in a \(O(log(F*S))\) asymptotic time complexity for the addition of a new spatial filter. \(F\) denotes the number of spatial filters (i.e., subscriptions) and \(S\) denotes the average filter size (i.e., number of \emph{peid}s associated with a spatial filter). 

The main memory mapping consists of a hash map with the \emph{peid}s as keys and hash sets with associated subscription identifiers as values. Here, the time complexity for addition of a subscription is \(O(S)\).

\subsection {Publication Filtering \& Routing}


Filtering publications is carried out in three distinct steps.
All active subscription identifiers are first filtered against an incoming publication by topic such as in traditional topic-based pub/sub with the topic filter mapping, resulting in a subset of all subscriptions denoted by $Sub_{tfiltered}$. 
Next, this subset is filtered by spatiality with the spatial filter mapping. In this process, we perform a lookup of the spatial filter map with the publication's \emph{peid}, resulting in the set of matching subscriptions denoted by $Sub_{matched}$. The function for this can be defined as follows: 
\begin{equation}
\begin{split}
    FilterLookup: peid \to \{ & sub_{1}, sub_{2}, \dots , sub_{n} \\
    & \mid sub_i \in S_{tfiltered} \}
\end{split}
\end{equation}
\vspace{-0.05in}

For the final step, each subscription of $Sub_{matched}$ is chosen for routing and the publication is added to the subscription's corresponding message queue, from which it is eventually delivered to the subscriber. 

Of course, the overhead of filtering is different for the two aforementioned modes. The lookup of the in-database mode has a complexity of \(O(log(F*S))\) and the main memory mode has a \(O(1)\) complexity.
Note that this architecture favors publication filtering over filter addition for the main memory mapping. An inverted index with the subscription identifiers as key and \emph{peid}s as values could provide an \(O(F)\) complexity for filtering and \(O(1)\) for filter addition.

\section{Implementation}
With the architecture from the preceding section, we end up with four distinct parts requiring implementation:

\begin{enumerate}
    \item A \emph{spatial database} that contains the geometric world model, providing neighborhood resolution as a service and storage for an in-database mapping of the results;
    \item A \emph{topic-based pub/sub broker} that incorporates spatial filtering capabilities based on the spatial filter mapping, which it constructs through the spatial database in response to neighborhood subscriptions;
    \item A \emph{message codec} that efficiently abstracts topic-based pub/sub communication while supporting representations for neighborhood descriptors in subscription messages and physical entity identifiers in publication messages;
    \item A \emph{spatial client} (library) that leverages the message codec and provides an interface to the spatial pub/sub middleware, enabling IoT and CPS devices to participate as subscribers and publishers.
\end{enumerate}

We based the realization of this architecture, which is detailed in the following sections, on several well-established open-source projects. The message broker \textit{Apache ActiveMQ Artemis}, together with enhancements to it, serves as the spatial pub/sub broker. The spatial database component is implemented using the PostGIS database management system. Location filter information is encoded using version 5 of the MQTT topic-based pub/sub protocol. Finally, the spatial client is built on top of Eclipse’s paho-mqtt project for MQTT client implementations.
More details on the implementation can be found in App.~\ref{details-implementation}.

\section{Evaluation Results} \label{sec:evaluations}

In this section, we show the behavior of the in-database and main memory mapping modes of \sysname.

\begin{figure*}[t!]
    \centering
    \subfloat[Throughput\label{fig:throughput}]{
        \includegraphics[width=0.31\textwidth]{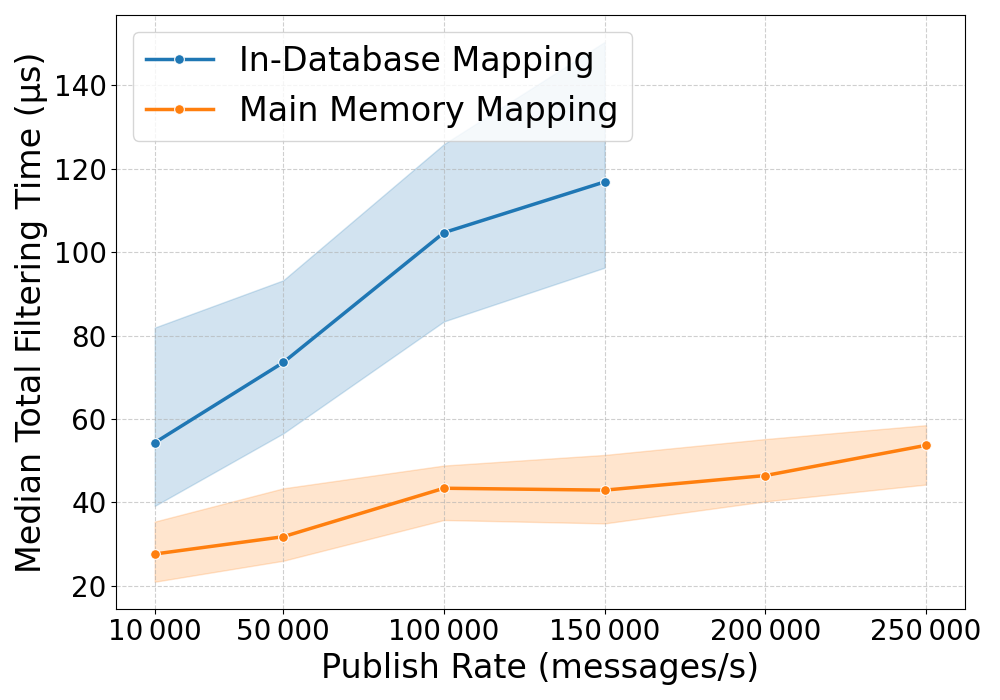}
    }
    \hfill
    \subfloat[Total Filtering Time\label{fig:num_subs_filtering_time}]{
        \includegraphics[width=0.31\textwidth]{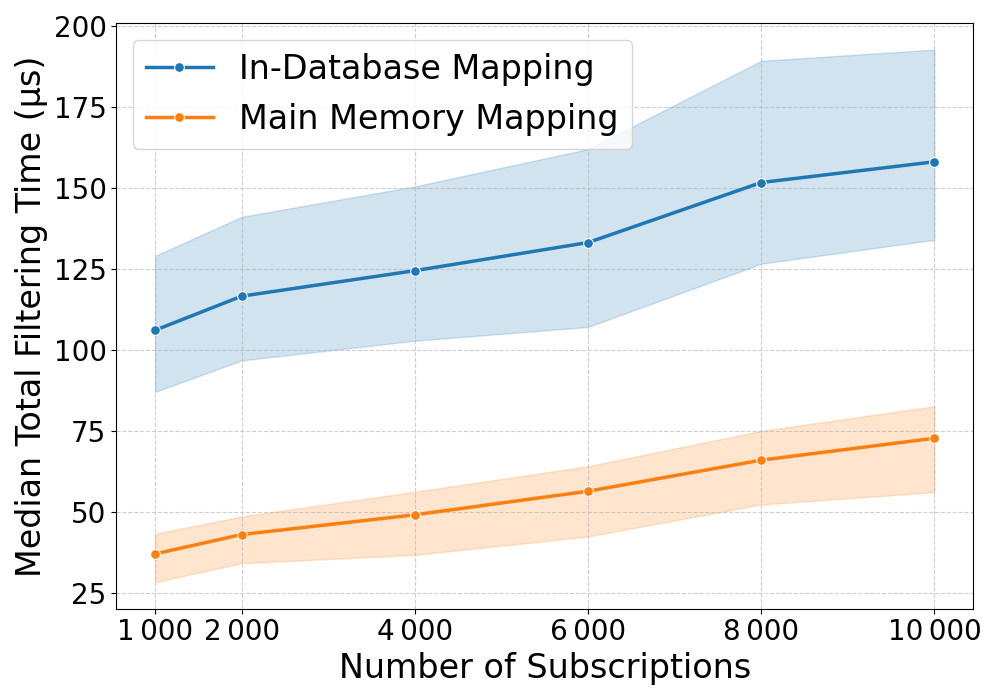}
    }
    \hfill
    \subfloat[Subscription Addition Time\label{fig:num_subs_add_filter_time}]{
        \includegraphics[width=0.31\textwidth]{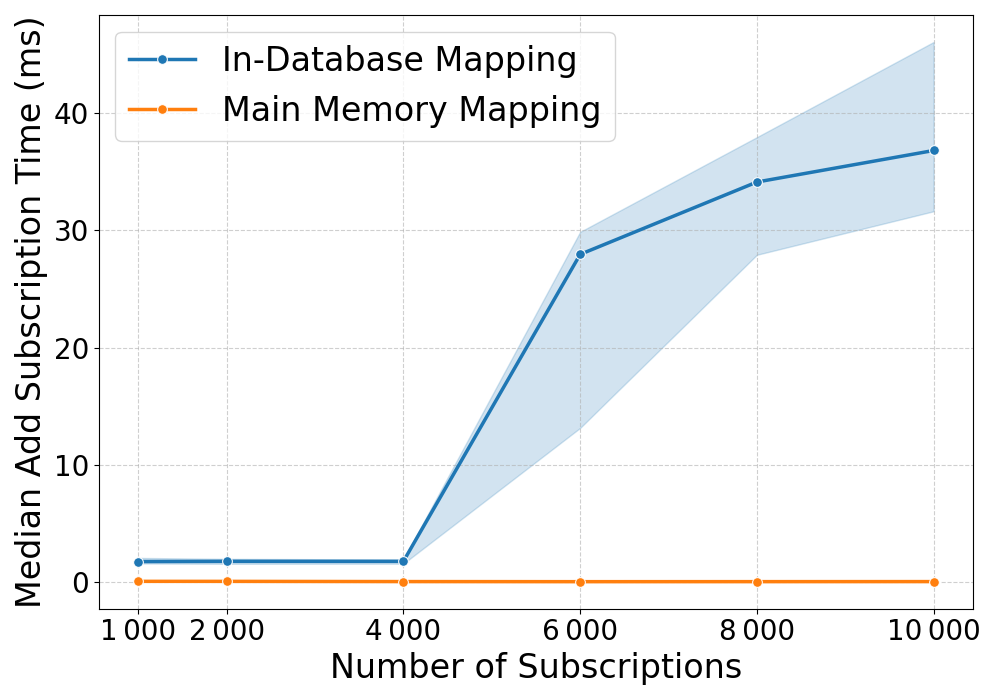}
    }
    \par\medskip  
    \subfloat[Total Filtering Time\label{fig:filter_size_filtering_time}]{
        \includegraphics[width=0.31\textwidth]{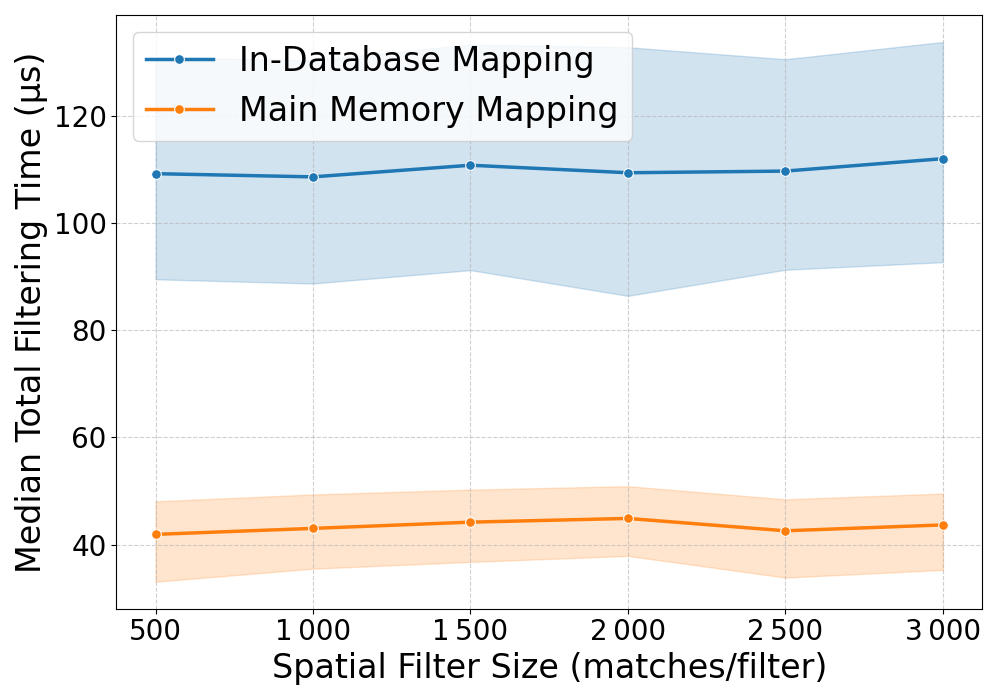}
    }
    \hfill
    \subfloat[Subscription Addition Time\label{fig:filter_size_add_filter_time}]{
        \includegraphics[width=0.31\textwidth]{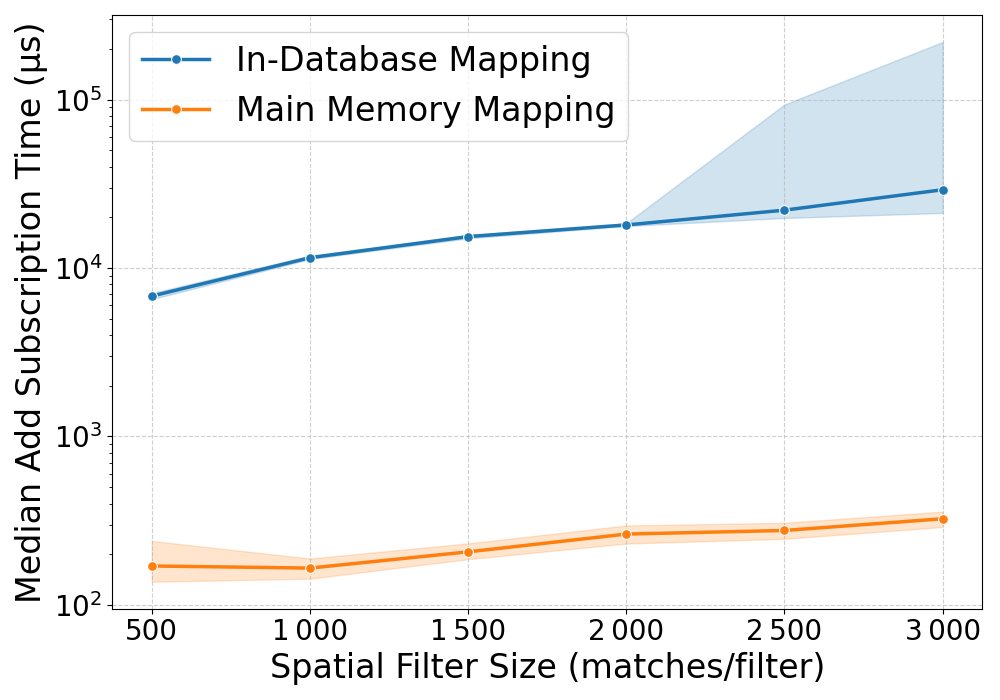}
    }
    \hfill
    \subfloat[Total Filtering Time: Real-World Data\label{fig:churn_rate_filter_time}]{
        \includegraphics[width=0.31\textwidth]{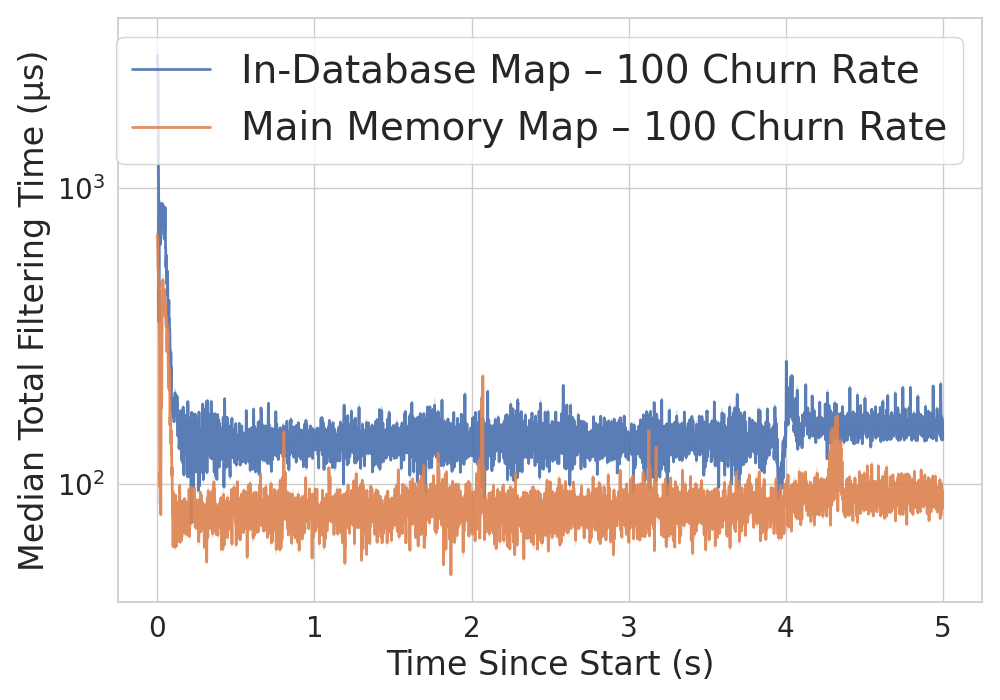}
    }
    
    
    \caption{Evaluation Results}
    \label{fig:random_comparison}
\end{figure*}

\subsection{Experimental Setup}
Our experimental setup consists of three identical machines with 13th Gen Intel Core i9-13900K 24-core processors.
Machine 1 runs the publishers, Machine 2 the subscribers, and Machine 3 the spatial broker. Furthermore, we deployed the spatial broker and database in separate Docker containers on Machine 3. With regard to the enhanced Artemis, we configured it according to the documentation's notes on performance tuning and turned persistence off. 

To establish the general performance of \sysname, we used synthetically generated spatial filter mappings as this provides better control of environmental variables like the matching of publications with subscriptions and the client distribution. The general characteristics of the spatial filter in terms of number of subscriptions as well as average filter size is also independent of neighborhood resolution itself. 
Moreover, we performed experiments on real data by leveraging Berlin OpenStreetMap data. The dataset is freely available through Geofabrik and ODIS Berlin and - after pre-processing - includes the road network as well as certain shops and businesses as a basis for neighborhoods. Subscribers are randomly generated on this by the road network's traffic distribution and assumed to be vehicles. These subscribers use a Zipfian distribution for randomly selecting reference points around them as the basis for neighborhoods. 

\subsection{Performance \& Scalability}
\subsubsection{Throughput}
Figure~\ref{fig:throughput} shows the throughput under a static, synthetic regime of 1\,000 subscriptions with a spatial filter size of 2\,000 \emph{peid}s by varying the publish rate for both types of mapping (in-database and main memory). The subscriptions are composed of a 100 subscribers with each having 10 subscriptions.
The graphs show the median total filtering time in $\mu$s as the throughput measure.
As expected, the main memory mode performs better than the in-database mode as the communication overhead between Artemis and PostGIS is not required in the main memory mode. In both cases, the filtering time increases with the publish rate, but the increase is steeper for the in-database mode. Consequently, the in-database mode performs well up to around 150\,000 messages/s, while the main memory mode scales much better to around 250\,000 messages/s. Above these the system ceases to function timely. 
Overall, considering the large scale of these experiments, w.r.t. number of subscriptions and spatial filter size, \sysname~exhibits high efficiency in terms of very low latencies for filtering in both modes (always in $\mu$s) and very high publish rates for especially main memory mode.

\begin{figure}[h!]
    \centering
    \includegraphics[width=0.75\linewidth]{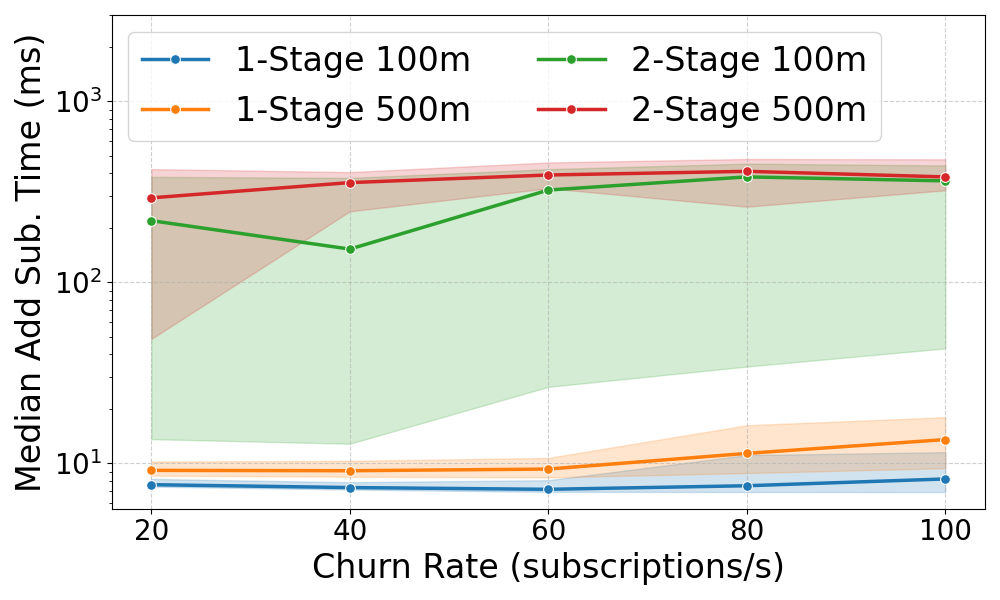}
    \caption{Subscription Addition Time: Real-World Data}
    \label{fig:churn_rate_add_filter_time}
\vspace{-8pt}
\end{figure}




\subsubsection{Impact of Number of Subscriptions}
The number of subscriptions can impact the total filtering time as discussed in Sec.~\ref{sec:middleware}. Figure~\ref{fig:num_subs_filtering_time} shows the total filtering time under a constant publish rate of 100\,000 messages/s and a fixed spatial filter size. Both mappings display a slowly growing filtering time with increasing number of subscriptions in the lower microseconds. While the in-database mode has a slightly faster growing rate and more variability, it has a consistently higher filtering time compared to the main memory mode. This is because of the additional overhead of database communication for the in-database mode. 
As can be seen, the filtering time of both modes is in $\mu$s, showing the filtering efficiency of \sysname.

Figure~\ref{fig:num_subs_add_filter_time} shows the time for the addition of a new filter (in $m$s) to the mappings under the same constant publish rate and fixed spatial filter size. To measure the filter addition time, we introduced a subscription churn rate (i.e., the number of subscriptions added and removed per second) of 10 to the system. The graph clearly shows the advantage of the main memory mode over the in-database mode by a constant addition time to a logarithmically increasing time. 

\subsubsection{Impact of Spatial Filter Size}
A second important factor on the behavior of the spatial filter is the average spatial filter size (no. of peids associated with a spatial filter). Figure~\ref{fig:filter_size_filtering_time} and Figure~\ref{fig:filter_size_add_filter_time} show the impact of a variable spatial filter size on the total filtering time and subscription addition time, respectively. The number of subscriptions is fixed at 1\,000 and the publish rate at 100\,000 messages per second. In the experiments in Figure~\ref{fig:filter_size_add_filter_time}, we introduce a subscription churn rate of 10 again to measure the filter addition time.  

Considering the filtering time in Figure~\ref{fig:filter_size_filtering_time}, both modes exhibit a constant time in the low $\mu$s over the chosen range of spatial filter size. The in-database mode's time is about 70 $\mu$s higher and shows a slightly increased variability. A logarithmic increase in this mode is not distinguishable under the chosen parameters, highlighting the efficiency of the system. 

For the filter addition time, the in-database mode shows a logarithmic increase and the main memory mode a more linear increase (note the log scale). The in-database mode performs worse for filter addition under this variable than the main memory mode, which exhibits values in the $\mu$s in contrast to the in-database mode that already starts out in the $m$s. 

\subsection{Performance under Real-World Data}
The final two experiments that Figure~\ref{fig:churn_rate_filter_time} and Figure~\ref{fig:churn_rate_add_filter_time} show use the above discussed real-world dataset. 
In Figure~\ref{fig:churn_rate_filter_time}, we conducted the experiments with 1\,000 1-stage neighborhood subscriptions. These resulted in a set of 592 publishers that had an overall publish rate of 50\,000 messages/s. 
This graph uses the data of the measurements taken at a churn rate of 100 and shows the rolling median total filtering time (in $\mu$s) over the first 5 seconds for both modes. After an initial warm-up period of around 200 $m$s both modes settle around a constant filtering time. Again, the main memory mode exhibits a better performance than the in-database mode. 

In Figure~\ref{fig:churn_rate_add_filter_time} we evaluate the behavior of multi-stage neighborhood resolution compared to single-stage neighborhoods for the main memory mode. Note that, in the case of real-world data, the time for actually resolving a neighborhood by the DBMS leads to additional overhead. Here we show the median subscription addition time under variable churn rate with regards to the scope of neighborhoods. The smaller neighborhood is depicted as 100m and the larger one is depicted as 500m. As expected, the single-stage resolution is faster than the multi-stage resolution, whereas the neighborhood scope only has a small effect in this case. The in-database mode showed similar behavior and was therefore omitted from the plot for clarity.

\section{Conclusion} \label{sec:conclusion}

In this paper, we presented \sysname, a pub/sub middleware that incorporates a novel approach of spatial awareness into subscriptions and publications. Subscriptions' spatial interests are defined through so-called neighborhoods
and resolved into sets of physical entity identifiers. These identifiers are associated with spatial attributes of a geometric world model and are added to publications to enable spatial filtering. 
Our evaluation results of a prototype implementation of \sysname~show highly efficient behavior of the system on synthetic as well as real-world data. Additionally, the results highlight the practical feasibility of incorporating this approach into existing pub/sub solutions and enabling new applications that rely on IoT communication based on the spatial relations of devices and loose coupling. 

\bibliography{library}
\bibliographystyle{plain}

\appendices

\section{Implementation Details} \label{details-implementation}
We based this implementation on several well-established open-source projects. The message broker \textit{Apache ActiveMQ Artemis}, together with enhancements to it, serves as the spatial pub/sub broker. The spatial database component is implemented using the \textit{PostGIS} database management system. Spatial filter information is encoded using MQTT version 5 topic-based pub/sub protocol. Finally, the spatial client is built on top of Eclipse’s \textit{paho-mqtt} project for MQTT client implementations.

Apache’s ActiveMQ Artemis is a multithreaded pub/sub broker and the direct successor to ActiveMQ Classic. It offers high throughput and efficiency as well as support for multiple pub/sub protocols, including MQTT and STOMP. In addition, the broker provides clustering capabilities for load distribution and persistence. Our enhancements extend Artemis with changes to the broker state and the introduction of a main-memory spatial filter mapping. 

PostGIS is an extension to the free and open-source relational DBMS PostgreSQL that provides comprehensive geospatial support. It offers spatial data types such as lines and polygons for both two- and three-dimensional space, spatial indexing, a wide range of spatial functions, and more. Access to the stored data from external applications is provided through JDBC. 

The transmittance of spatial publications and neighborhood subscriptions requires a clear and well-defined representation of semantics that all communication participants understand. We chose the topic-based MQTT version 5 (MQTT5) pub/sub protocol as the basis, as it is the de-facto industry standard, already provides the general semantics needed and supports application-level customization. Consequently, our implementation conforms fully to the existing standard. 

MQTT5 defines 15 different control packet types for conveying information such as publications. In addition, it supports three levels of quality of service (QoS) for message delivery guarantees, denoted 0, 1, and 2. QoS 0 provides at-most-once semantics, QoS 1 at-least-once semantics, and QoS 2 exactly-once semantics. The focus of this work limits itself to QoS 0 delivery semantics and does not study the implications of client authentication, which MQTT5 also supports. With these constraints, the number of relevant control packet types is reduced to the five basic ones: \emph{Connect}, \emph{Disconnect}, \emph{Subscribe}, \emph{Unsubscribe}, and \emph{Publish}. 

Some MQTT5 control packet types support fields referred to as user properties, which are key–value pairs encoded as UTF-8 strings and can be used freely to provide application-specific information without any changes to the standardized protocol itself. Older versions of MQTT do not support user properties. The Connect, Disconnect, Subscribe, Unsubscribe, and Publish packet types all support user properties. 
For a neighborhood subscribe, the Subscribe packet type requires two user properties: a neighborhood subscription identifier and a neighborhood descriptor that encodes the neighborhood descriptor as a JSON UTF-8 String. The Unsubscribe control packet requires only the neighborhood subscription identifier. (Note: the identifiers could also be defined by the broker and returned in the \textit{Suback} control packet type, which acknowledges a subscription. This could guarantee the system-wide uniqueness of identifiers.)

For the Publish control packet, two cases must be distinguished: publications sent from a client to the broker and filtered publications forwarded from the broker to subscribers. In the first case, the Publish packet includes a user property containing only the physical entity identifier. In the second case, the broker injects the matching spatial filter identifier into the publication as an additional user property so that the subscriber can associate the publication with a specific neighborhood subscription. Without this additional property, a client with two neighborhood subscriptions to the same topic could not determine from which neighborhood a given publication originates.


Clients — acting as publishers and subscribers — that utilize the middleware must, in principle, only adhere to the MQTT5 specification for communication and encode the additional spatial information using user properties. The actual client applications themselves are irrelevant to the middleware as long as the world model is maintained within the spatial database. Consequently, any MQTT5-compliant client library is sufficient, provided it supports writing and reading user properties.

\textit{Paho}, an IoT project of the Eclipse Foundation, offers MQTT client implementations in various programming languages. Its Java, Python, C, and C++ libraries support MQTT5 in addition to MQTT versions 3.1 and 3.1.1. For our implementation, we selected the \textit{paho-mqtt} Python library, as it provides a simple interface and enjoys broad adoption. A lightweight wrapper around the library’s \texttt{MQTTClient} object, as well as its \texttt{subscribe}, \texttt{unsubscribe}, and \texttt{publish} methods, is sufficient to enable communication with the spatial pub-sub middleware. The \texttt{unsubscribe} and \texttt{publish} methods include an additional integer argument for the neighborhood subscription identifier and the physical entity identifier, while the \texttt{subscribe} method takes, as an additional argument, a string representing the neighborhood descriptor in JSON format.

\end{document}